\documentclass[12pt,a4j]{article}
\usepackage[dvips]{graphicx}
\usepackage{amsmath}
\usepackage{amssymb}
\usepackage{amsfonts}
\usepackage{bm}
\usepackage{enumerate}

\begin{document}
\baselineskip=6mm
\centerline{\bf Multisoliton formulas for the multi-component Fokas-Lenells equation} \par
\centerline{\bf with nonzero boundary conditions}\par
\bigskip
\centerline{Yoshimasa Matsuno}\par
\centerline{Division of Applied Mathematical Science}\par
\centerline{ Graduate School of Sciences and Technology for Innovation} \par
\centerline{Yamaguchi University}\par
\centerline{E-mail address: matsuno@yamaguchi-u.ac.jp}\par

\bigskip
\noindent {\bf Abstract}.  The multi-component Fokas-Lenells  equation is considered. In particular, we present the multisoliton formulas for the system
with plane-wave boundary conditions, as well as with mixed zero and plane-wave boundary conditions.  A direct approach is employed to construct solutions,
showing that for both boundary conditions, the multisoliton solutions have compact determinantal expressions.
 \par
\bigskip
\bigskip
\noindent {\bf 1. Introduction}\par
\medskip
The Lax pair of the integrable multi-component Fokas-Lenells system is given by 
$$\Psi_x=U\Psi, \quad \Psi_t=V\Psi,  \eqno(1.1a)$$
$$U=\begin{pmatrix} {{\rm i}\over 2}\zeta^2 & -{\rm i}\zeta{\bf u}_x\\ {\rm i}\zeta{\bf v}_x^T & -{{\rm i}\over 2}\zeta^2I\end{pmatrix}=(u_{jk}), \quad
V=\begin{pmatrix}-{\rm i\over 2\zeta^2} -{\rm i}{\bf u}{\bf v}^T & {1\over\zeta}{\bf u}\\ {1\over\zeta}{\bf v}^T & 
      {\rm i\over 2\zeta^2}\ I +{\rm i}{\bf v}^T{\bf u}\end{pmatrix}=(v_{jk}), \eqno(1.1b)$$
      where $\zeta$ is the spectral parameter, and 
     $ {\bf u}=(u_1, u_2, ..., u_n)$ and $ {\bf v}=(v_1, v_2, ..., v_n),  {\bf u},{\bf v} \in \mathbb{C}^n$ are $n$-component row vectors.
      \par
      It follows from the compatibility condition of the Lax pair that 
      $U_t-V_x+UV-VU=O$. This yields the system of nonlinear PDEs for ${\bf u}$ and ${\bf v}$: 
      $${\bf u}_{xt}-{\bf u}+{\rm i}({\bf u}_x{\bf v}^T{\bf u}+{\bf u}{\bf v}^T{\bf u}_x)={\bf 0}, \eqno(1.2a)$$
      $${\bf v}_{xt}-{\bf v}-{\rm i}({\bf v}_x{\bf u}^T{\bf v}+{\bf v}{\bf u}^T{\bf v}_x)={\bf 0}. \eqno(1.2b)$$
      \par
 The system  (1.2) can be reduced from the first negative flow of the matrix derivative NLS hierarchy [1-4]. 
There arise several integrable PDEs from the reductions of the system (1.2).  Specifically, if we put  $v_j=\sigma_ju_j^*,\ \sigma_j=\pm 1\ (j=1, 2, ..., n)$, then the above system reduces to
$$u_{j, xt}=u_j-{\rm i}\left\{\left(\sum_{k=1}^n\sigma_ku_{k,x}u_k^*\right)u_j+\left(\sum_{k=1}^n\sigma_ku_ku_k^*\right)u_{j,x}\right\}, \ (j=1, 2, ..., n). \eqno(1.3)$$
The system of PDEs (1.3) is the basic equation that we consider here. The two special cases reducing from the system (1.3) are  particularly important: \par
\medskip
\noindent 1) ${n=1}$: FL equation [5, 6] 
$$u_{xt}=u-2{\rm i}\sigma|u|^2u_x, \ (u\equiv u_1, \sigma_1= \sigma). \eqno(1.4) $$
\noindent 2) ${n=2}$: two-component FL system [4, 7]
$$u_{1,xt}=u_1-{\rm i}\left\{(2|u_1|^2+\sigma|u_2|^2)u_{1,x}+{\rm i}\sigma u_1u_2^*u_{2,x}\right\}, \eqno(1.5a)$$
$$u_{2,xt}=u_2-{\rm i}\left\{(|u_1|^2+2\sigma|u_2|^2)u_{2,x}+{\rm i}\sigma u_2u_1^*u_{1,x}\right\}, \eqno(1.5b)$$
\centerline{$(\sigma_1=1, \sigma_2=\sigma).$}
\par
\medskip
The $N$-soliton solutions of the FL equation have been constructed for both zero and plane-wave boundary conditions [8, 9] while
for the general $n$-component system, we have obtained the bright $N$-soliton solution with zero boundary conditions [10].
The purpose of the current work is to  present the $N$-soliton formulas of the system (1.3) with  the following two types of the boundary conditions: \par
\medskip
\noindent 1) Plane-wave boundary conditions \par
$$u_j \sim \rho_j\,{\rm exp}\,{\rm i}\left(k_jx-\omega_jt+\phi_j^{(\pm)}\right), \quad x\rightarrow \pm\infty, \quad (j=1, 2, ..., n), \eqno(1.6a)$$
with the linear dispersion relation 
$$k_j\omega_j=1+\sum_{s=1}^n\sigma_sk_s\rho_s^2+\sum_{s=1}^n\sigma_s\rho_s^2k_j, \quad (j=1, 2, ..., n). \eqno(1.6b)$$
\medskip
\noindent 2) Mixed type boundary conditions \par
$$u_j \sim 0, \quad x \rightarrow \pm\infty, \quad (j=1, 2, ..., m), \eqno(1.7a)$$
$$u_{m+j} \sim \rho_j\,{\rm exp}\,{\rm i}\left(k_jx-\omega_jt+\phi_j^{(\pm)}\right), \quad x\rightarrow \pm\infty, \quad (j=1, 2, ...,n- m), \eqno(1.7b)$$
with the linear dispersion relation 
$$k_j\omega_j=1+\sum_{s=1}^{n-m}\sigma_sk_s\rho_s^2+\sum_{s=1}^{n-m}\sigma_s\rho_s^2k_j, \quad (j=1, 2, ...,n- m). \eqno(1.7c)$$
\par
In this short note, we provide the main results only, and the details will be reported elsewhere. \par
\bigskip
\leftline{ \bf 2. The  $N$-soliton formula with plane-wave boundary conditions}\par
\noindent{2.1. Bilinearization} \par
\medskip
Here, we present the multisoliton solutions of the system (1.3) with plane-wave boundary conditions (1.6).  The direct approach is used to
obtain solutions. To this end, we start from the following proposition: \par
\medskip
\noindent{\bf Proposition 1}. 
{\it Under the dependent variable transformations
$$u_j=\rho_j{\rm e}^{{\rm i}(k_jx-\omega_jt)}\,{g_j\over f}, \ (j=1, 2, ..., n), \eqno(2.1)$$
the multi-component FL system (1.3) can be decoupled into the system of equations
$$D_tf\cdot f^*={\rm i}\sum_{s=1}^n\sigma_s\rho_s^2(g_sg_s^*-ff^*), \eqno(2.2a)$$
$$D_xD_tf\cdot f^*-{\rm i}\sum_{s=1}^n\sigma_s\rho_s^2D_xg_s\cdot g_s^*+{\rm i}\sum_{s=1}^n\sigma_s\rho_s^2D_xf\cdot f^*+2\sum_{s=1}^n\sigma_sk_s\rho_s^2(g_sg_s^*-ff^*)=0, \eqno(2.2b)$$
$$f^*\left[g_{j,xt}f-(f_x-{\rm i}k_jf)g_{j,t}-{\rm i}{1\over k_j}\left(1+\sum_{s=1}^n\sigma_sk_s\rho_s^2\right)D_xg_j\cdot f\right]$$
$$=f_t^*(g_{j,x}f-g_jf_x+{\rm i}k_jg_jf), \ (j=1,2, ..., n), \eqno(2.2c)$$
where $f=f(x, t)$ and $g_j=g_j(x, t)$ are the complexed-valued functions of $x$ and $t$, and the bilinear operators $D_x$ and $D_t$ are defined by
$$D_x^mD_t^nf\cdot g=\left({\partial\over\partial x}-{\partial\over\partial x^\prime}\right)^m
\left({\partial\over\partial t}-{\partial\over\partial t^\prime}\right)^n
f(x, t)g(x^\prime,t^\prime)\Big|_{ x^\prime=x,\,t^\prime=t}$$
with $m$ and $n$ being nonnegative integers.} \par
\medskip
\noindent  {\bf Remark 1}.\par
\medskip
\noindent 1) We can decouple the last equation into a system of bilinear equations
$$g_{j,xt}f-(f_x-{\rm i}k_jf)g_{j,t}-{\rm i}{1\over k_j}\left(1+\sum_{s=1}^n\sigma_sk_s\rho_s^2\right)D_xg_j\cdot f=h_jf_t^*, \eqno(2.3a)$$
$$g_{j,x}f-g_jf_x+{\rm i}k_jfg_j=h_jf^*, \eqno(2.3b)$$
where $h_j=h_j(x, t)$ are the complexed-valued functions of $x$ and $t$. \par
\medskip
\noindent 2) If we introduce the variables $q_j=u_{j,x}$, then
$$q_j=\left(\rho_j{\rm e}^{{\rm i}(k_jx-\hat\omega_jt)}\,{g_j\over f}\right)_x=\rho_j{\rm e}^{{\rm i}(k_jx-\hat\omega_jt)}{h_jf^*\over f^2},
\quad \hat\omega_j=k_j^2+2\sum_{s=1}^n\sigma_s\rho_s^2k_j,\quad (j=1, 2, ..., n), \eqno(2.4)$$
 solve the $n$-component derivative NLS system
$${\rm i}q_{j,t}+q_{j,xx}+2{\rm i}\left[\left(\sum_{s=1}^n\sigma_s|q_s|^2\right)q_j\right]_x=0, \quad (j=1, 2, ..., n). \eqno(2.5)$$
\noindent{ 2.2. $N$-soliton solution}\par
\medskip
\noindent {\bf Theorem 1}.
{\it The $N$-soliton solution of the system of bilinear equations (2.2) is given in terms of the following determinants.
$$f=|D|,\quad g_s=|G_s|, \quad (s=1, 2, ..., n), \eqno(2.6a)$$
$$D=(d_{jk})_{1\leq j, k\leq N}, \quad d_{jk}=\delta_{jk}-{{\rm i}p_j\over p_j+p_k^*}\,z_jz_k^*, \eqno(2.6b)$$
$$G_s=(g^{(s)}_{jk})_{1\leq j, k\leq N}, \quad g^{(s)}_{jk}=\delta_{jk}-{{\rm i}p_k^*\over p_j+p_k^*}{p_j-{\rm i}k_s\over p_k^*+{\rm i}k_s}\,z_jz_k^*, \eqno(2.6c)$$
$$z_j={\rm exp}\left[p_jx+{1\over p_j}\left(1+\sum_{s=1}^n\sigma_sk_s\rho_s^2\right)t+\zeta_{j0}\right], \quad (j=1, 2, ..., N). \eqno(2.6d)$$
Here, $p_j$ and $\zeta_{j0}\ (j=1, 2, ..., N)$  are arbitrary complex parameters. The former parameters are impozed on $N$ constraints
$$\sum_{s=1}^n\sigma_s(k_s\rho_s)^2\,{{\rm i}(p_j-p_j^*)+k_s\over (p_j-{\rm i}k_s)(p_j^*+{\rm i}k_s)}=-1, \quad \ (j=1, 2, ..., N). \eqno(2.7)$$ }\par
The expressions (2.1) with the tau-functions (2.6) give the dark soliton solutions with plane-wave boundary conditions. 
The analysis of the one-component system (i.e., FL equation) has been performed in [9] where the detailed description of the dark soliton solutions has been given. 
\par
\medskip
\noindent {\bf Remark 2}. \par
\medskip
\noindent 1) The proof of the $N$-soliton solution  can be done by means of an elementary calculation using the basic formulas of determinants, i.e., 
$${\partial\over\partial x}|D|=\sum_{j,k=1}^N{\partial d_{jk}\over\partial x}D_{jk},\quad (D_{jk}: {\rm cofactor\ of}\ d_{jk}),$$
$$\begin{vmatrix} D & {\bf a}^T\\ {\bf b} & z\end{vmatrix}=|D|z-\sum_{j,k=1}^ND_{jk}a_jb_k,$$
$$|D({\bf a}, {\bf b}; {\bf c}, {\bf d})||D|= |D({\bf a}; {\bf c})||D({\bf b}; {\bf d})|-|D({\bf a}; {\bf d})||D({\bf b}; {\bf c})|,\ ({\rm Jacobi's\ identity}),$$
with the notation
$$\begin{vmatrix} D & {\bf b}^T\\ {\bf a} & 0\end{vmatrix}=|D({\bf a}; {\bf b})|,\quad \begin{vmatrix} D& {\bf c}^T &{\bf d}^T\\ {\bf a}&0&0\\{\bf b}&0&0\end{vmatrix}=|D({\bf a}, {\bf b}; {\bf c}, {\bf d})|.$$
\medskip
\noindent 2)  The tau-functions $h_s$ are given by
$$h_s={\rm i}k_s|H_s|, \quad  H_s=(h^{(s)}_{jk})_{1\leq j, k\leq N},\quad h^{(s)}_{jk}=\delta_{jk}+{{\rm i}p_j\over p_j+p_k^*}{p_j-{\rm i}k_s\over p_k^*+{\rm i}k_s}\,z_jz_k^*.$$
\noindent{2.3. Derivation of constraints (2.7)} \par
\medskip
In the case of plane-wave boundary conditions, the $n$ constraints must be imposed among the complex parameters $p_j\ (j=1, 2, ..., N)$. We derive these constraints from the
Lax pair (1.1) of the system.
The spatial part of the Lax pair with seed solutions 
 $$u_j=\rho_j{\rm e}^{{\rm i}\theta_j}, \quad \theta_j=k_jx-\omega_jt, \quad (j=1, 2, ..., n),$$ 
  are given by
 $$\Psi_x=U\Psi, \quad 
  U=\begin{pmatrix} {{\rm i}\over 2}\zeta^2 & k_1\rho_1\zeta {\rm e}^{{\rm i}\theta_1} & \cdots & k_n\rho_n\zeta {\rm e}^{{\rm i}\theta_n} \\
      \sigma_1k_1\rho_1\zeta {\rm e}^{-{\rm i}\theta_1}& -{{\rm i}\over 2}\zeta^2 & \cdots & 0\\                     
      \vdots & \vdots& \ddots & \vdots \\
      \sigma_nk_n\rho_n\zeta {\rm e}^{-{\rm i}\theta_n}& 0 & \cdots & -{{\rm i}\over 2}\zeta^2 \end{pmatrix}. \eqno(2.8)$$
 Introduce a new wavefunction $\Psi_0$ by $\Psi=P\Psi_0$, where $P$ is a diagonal matrix $P={\rm diag}(1, {\rm e}^{{\rm i}\theta_1}, ...,, {\rm e}^{{\rm i}\theta_n})$.
 Then, $\Psi_0$ satisfies the matrix equation
 $$\Psi_{0, x}=(P_xP^{-1}+PUP^{-1})\Psi_0\equiv U_0\Psi_0, \quad 
 U_0=\begin{pmatrix} {{\rm i}\over 2}\zeta^2 & k_1\rho_1\zeta  & \cdots & k_n\rho_n\zeta  \\
      \sigma_1k_1\rho_1\zeta & {\rm i}k_1-{{\rm i}\over 2}\zeta^2 & \cdots & 0\\                     
      \vdots & \vdots& \ddots & \vdots \\
      \sigma_nk_n\rho_n\zeta & 0 & \cdots & {\rm i}k_n-{{\rm i}\over 2}\zeta^2 \end{pmatrix}. \eqno(2.9)$$
The characteristic equation of $U_0$ reads
$|U_0-I_{n+1}\mu|=0 $, i.e.,
$$\begin{vmatrix} {{\rm i}\over 2}\zeta^2-\mu & k_1\rho_1\zeta  & \cdots & k_n\rho_n\zeta  \\
      \sigma_1k_1\rho_1\zeta & {\rm i}k_1-{{\rm i}\over 2}\zeta^2-\mu & \cdots & 0\\                     
      \vdots & \vdots& \ddots & \vdots \\
      \sigma_nk_n\rho_n\zeta & 0 & \cdots & {\rm i}k_n-{{\rm i}\over 2}\zeta^2-\mu \end{vmatrix}=0. \eqno(2.10)$$
 Expanding the above determinant in $\mu$ yields
 $${{\rm i}\over 2}\zeta^2-\mu=-\zeta^2\sum_{s=1}^n{\sigma_s(k_s\rho_s)^2\over \mu+{{\rm i}\over 2}\zeta^2-{\rm i}k_s}. \eqno(2.11)$$
 Let $\mu+{{\rm i}\over 2}\zeta^2=p$ and assume $\zeta^2$ be real and $p$ be complex. Then
 $${\rm i}\zeta^2-p=-\zeta^2\sum_{s=1}^n{\sigma_s(k_s\rho_s)^2\over p-{\rm i}k_s}, \quad -{\rm i}\zeta^2-p^*=-\zeta^2\sum_{s=1}^n{\sigma_s(k_s\rho_s)^2\over p^*+{\rm i}k_s}. \eqno(2.12)$$
 It follows from the above two relations that
 $$\sum_{s=1}^n\sigma_s(k_s\rho_s)^2\,{{\rm i}(p-p^*)+k_s\over (p-{\rm i}k_s)(p^*+{\rm i}k_s)}=-1, \eqno(2.13)$$
which yields (2.7) upon putting $p=p_j$. \par
\bigskip
\noindent{ \bf 3. The  $N$-soliton formula with mixed type boundary conditions}\par
\noindent{3.1. Bilinearization} \par
\medskip
The bilinearization of the system (1.3) with mixed type boundary conditions (1.7) can be performed by the following proposition.@\par
\medskip
\noindent {\bf Proposition 2}. 
{\it Under the dependent variable transformations
$$u_j={\rm e}^{-{\rm i}\hat\lambda t}\,{h_j\over f}, \quad \left(j=1, 2, ..., m,\ \hat\lambda=\sum_{s=1}^{n-m}\sigma_{m+s}\rho_s^2\right), \eqno(3.1a)$$
$$u_{m+j}=\rho_j\,{\rm e}^{{\rm i}(k_jx-\omega_jt)}\,{g_j\over f}, \quad (j=1, 2, ..., n-m), \eqno(3.1b)$$
the multi-component FL system (1.3) can be decoupled into the system of equations
$$D_tf\cdot f^*={\rm i}\sum_{s=1}^m\sigma_sh_sh_s^*+{\rm i}\sum_{s=1}^{n-m}\sigma_{m+s}\rho_s^2(g_sg_s^*-ff^*), \eqno(3.2a)$$
$$D_xD_tf\cdot f^*-{\rm i}\sum_{s=1}^m\sigma_sD_xh_s\cdot h_s^*-{\rm i}\sum_{s=1}^{n-m}\sigma_s\rho_s^2D_xg_s\cdot g_s^*
+{\rm i}\sum_{s=1}^{n-m}\sigma_{m+s}\rho_s^2D_xf\cdot f^*$$
$$+2\sum_{s=1}^{n-m}\sigma_sk_s\rho_s^2(g_sg_s^*-ff^*)=0, \eqno(3.2b)$$
$$f^*(h_{j,xx}f-h_{j,t}f_x-\lambda h_jf)=f_t^*(h_{j,x}f-h_jf_x), (j=1, 2, ..., m), \eqno(3.2c)$$
$$f^*\left\{g_{j,xt}f-(f_x-{\rm i}k_jf)g_{j,t}-{{\rm i}\lambda\over k_j}D_xg_j\cdot f\right\}
=f_t^*(g_{j,x}f-g_jf_x+{\rm i}k_jg_jf), \ (j=1, 2, ..., n-m), \eqno(3.2d)$$
where $\lambda=1+\sum_{s=1}^{n-m}\sigma_sk_s\rho_s^2$.}  \par
\medskip
\noindent{ 3.2. $N$-soliton solution}\par
\medskip
\noindent {\bf Theorem 2}.
{\it The $N$-soliton solution of the system of bilinear equations (3.2) is given in terms of the following determinants.
$$f=|D|,\quad D=(d_{jk})_{1\leq j, k\leq N}, \quad d_{jk}= {z_jz_k^*-{\rm i}p_k^*c_{jk}\over p_j+p_k^*}, \quad z_j={\rm exp}\left(p_jx+{\lambda\over p_j}\,t\right), \eqno(3.3a)$$
$$h_j=-{1\over\lambda}|D({\bf a}_j^*; {\bf z}_t)|,  \quad (j=1, 2, ..., m), \eqno(3.3b)$$
$$g_j=|D|+{{\rm i}\over\lambda}|D({\bf z}_j^*; {\bf z}_t)|, \quad (j=1, 2. ..., n-m), \eqno(3.3c)$$
$${\bf z}=(z_1, z_2, ..., z_N), \ {\bf z}_t=\left({\lambda\over p_1}\,z_1, {\lambda\over p_2}\,z_2, ..., {\lambda\over p_N}\,z_N\right), \eqno(3.3d)$$
$${\bf a}_j=(\alpha_{j1}, \alpha_{j2}, ..., \alpha_{jN}),\quad (j=1, 2, ..., m), \eqno(3.3e)$$
$$c_{jk}={\sum_{s=1}^m\sigma_s\alpha_{sj}\alpha_{sk}^* \over 1+\sum_{s=1}^{n-m}\sigma_s(k_s\rho_s)^2{{\rm i}(p_j-p_k^*)+k_s\over (p_j-{\rm i}k_s)(p_k^*+{\rm i}k_s)}}, \quad (j, k=1, 2, ..., N), \eqno(3.3f)$$
where $p_j\ (j=1, 2, ..., N)$ and $\alpha_{jk}\ (j=1, 2, ..., m; k=1, 2, ..., N)$ are arbitrary complex parameters.} \par
\medskip
The components from $(3.1a)$ take the form of the bright solitons with zero background whereas those of $(3.1b)$ represent the dark solitons with plane-wave background. 
The properties of the bright soliton solutions of the FL equation  have been explored in detail in [8].  It should be remarked that unlike  purely  plane-wave boundary conditions, no constraints
are imposed on the parameters $p_j$. Consequently,  the analysis of solutions becomes more easier than that of solutions for plane-wave boundary conditions. 
\par
\medskip
\noindent{\bf Remark 3}. \par
\medskip
\noindent 1) When compared with the soliton solutions  with the pure plane-wave boundary conditions, the parameters $p_j$ can be chosen arbitrary. Consequently, the explicit form
of the $N$-soliton solution is available without solving algebraic equations like (2.7).  \par
\medskip
\noindent 2)  If we put $\rho_j=0,\ (j=1, 2, ..., n-m)$, then $(3.1a)$ and (3.3) yield the bright $N$-soliton solution of the system (1.3) with the zero boundary conditions $u_j\rightarrow 0, |x|\rightarrow \infty$ [10].
\par
\bigskip
\noindent{\bf Acknowledgement} \par
\bigskip
This work was partially supported by the Research Institute for Mathematical Sciences, a Joint Usage/Research Center located in Kyoto University. \par
\bigskip
\noindent{\bf References} \par
\baselineskip=4mm
\begin{enumerate}[{[1]}]
\item A. P. Fordy, Derivative nonlinear Schr\"odinger equations and Hermitian symmetric spaces, {\it J. Phys. A: Math. Gen.} {\bf 17}  (1984) 1235-1245.
\item T. Tsuchida and M. Wadati, New integrable systems of derivative nonlinear Schr\"odinger equations with multiple components, {\it Phys. Lett.} A {\bf 257} (1999) 53-64.
\item T. Tsuchida, New reductions of integrable matrix partial differential equations: $Sp(m)$-invariant system, {\it J. Math. Phys.} {\bf 51} (2010) 053511.
\item B. Guo and L. Ling, Riemann-Hilbert approach and $N$-soliton formula for coupled derivative Schr\"odinger equation, {\it J. Math. Phys.} {\bf 53} (2012) 073506.
\item A. S. Fokas, On a class of physically important integrable equations, {\it Physica } D {\bf 87} (1995) 145-150.
\item J. Lenells, Exactly solvable model for nonlinear pulse propagation in optical fibers, {\it Stud. Appl. Math.} {\bf 123} (2009) 215-232.
\item L. Ling, B.-F. Feng and Z. Zhu, General  soliton solutions to a coupled Fokas-Lenells equation, {\it Nonlinear Anal.: Real World Applications} {\bf 40} (2018) 185-214.
\item Y. Matsuno, A direct method of solution for the Fokas-Lenells derivative nonlinear Schr\"odinger equation: I. Bright soliton solutions, {\it J. Phys. A: Math. Theor.} {\bf 45} (2012) 235202.
\item Y. Matsuno, A direct method of solution for the Fokas-Lenells derivative nonlinear Schr\"odinger equation: II. Dark soliton solutions, {\it J. Phys. A: Math. Theor.} {\bf 45} (2012) 475202.
\item Y. Matsuno,  Multi-component generalization of the Fokas-Lenells equation, RIMS K$\hat{\rm o}$ky$\hat {\rm u}$roku  {\bf 2076} (2018) 224-231.

\end{enumerate}

\end{document}